\documentclass[preprint,showpacs,preprintnumbers,aps,pra,floatfix]{revtex4-1}

\usepackage{bm}
\usepackage{graphicx}
\usepackage{amsmath}
\usepackage{amssymb}
\usepackage[usenames]{color}
\usepackage{subfigure}
\usepackage{amsfonts,amstext,graphics,graphicx}
\usepackage{color}

\graphicspath{{figs/}{figs_lo/}}

\newtheorem{theorem}{{\bf Theorem}}

\def\sgn{{\;{\rm sgn}}}
\def\<{{\langle}}
\def\>{{\rangle}}

\def\p{\partial}
 
\def\bq{\begin{equation}}
\def\eq{\end{equation}}
\def\bqy{\begin{eqnarray}}
\def\eqy{\end{eqnarray}}
\def\bqyn{\begin{eqnarray*}}
\def\eqyn{\end{eqnarray*}}

\def\calb{{\cal B}}

\def\calf{{\cal F}}

\def\calh{{\cal H}}

\def\ga{\gamma}
\def\Ga{\Gamma}
\def\si{\sigma}
\def\ze{\zeta}

\def\om{\omega}

\def\varep{\varepsilon}
\def\vep{\varepsilon}

\def\R{\mathbb{R}}

\def\N{\mathbb{N}}
\def\T{\mathbb{T}}

\def\A{\mathfrak{A}}
\def\J{\mathfrak{J}}
\def\T{\mathfrak{T}}
\def\E{\mathfrak{E}}
\def\L{\mathfrak{L}}

\newcommand{\comment}[1]{}
\newcommand{\mathnotation}[2]{\newcommand{#1}{\ensuremath{#2}}}

\mathnotation{\ldef}{\mathrel{\raisebox{.069ex}{:}\!\!=}}
\mathnotation{\rdef}{\mathrel{=\!\!\raisebox{.069ex}{:}}}	 
\mathnotation{\dint}{\,{\mathrm{d}}}		 
\mathnotation{\onehalf}{\tfrac{1}{2}}		 
\mathnotation{\qq}{\theta}

 \def\Xint#1{\mathchoice
{\XXint\displaystyle\textstyle{#1}}%
{\XXint\textstyle\scriptstyle{#1}}%
{\XXint\scriptstyle\scriptscriptstyle{#1}}%
{\XXint\scriptscriptstyle\scriptscriptstyle{#1}}%
\!\!\int}
\def\XXint#1#2#3{{\setbox0=\hbox{$#1{#2#3}{\int}$ }
\vcenter{\hbox{$#2#3$ }}\kern-.5\wd0}}

\def\dashint{\Xint-}

\begin{document}

\title{Continuum Hamiltonian Hopf Bifurcation II}
 \author{George I. Hagstrom}
\email{georgehagstrom@nyu.edu}
\affiliation{CIMS, New York University, New York, NY, USA}
\author{Philip J. Morrison}
\email{morrison@physics.utexas.edu}
\affiliation{Department of Physics \& Institute for Fusion Studies, University of Texas, Austin 78712-0262, USA}

\begin{abstract}

Building on the development of \cite{chaptI},  bifurcation of unstable modes that emerge 
from continuous spectra in a class of infinite-dimensional noncanonical Hamiltonian systems is investigated. 
Of main interest is a bifurcation termed the continuum Hamiltonian Hopf (CHH) bifurcation, 
which is an infinite-dimensional analog of the usual Hamiltonian Hopf (HH) bifurcation.  
Necessary notions pertaining to  spectra,  structural stability, signature of the continuous spectra, and 
normal forms are described.  The theory developed is applicable to a wide class of 2+1 noncanonical Hamiltonian matter models, but  the specific example of the Vlasov-Poisson system linearized about homogeneous 
(spatially independent) equilibria  is treated in detail. For this example,  structural (in)stability 
is established in an appropriate functional analytic setting, and two kinds of bifurcations are 
considered, one at infinite and one at finite wavenumber.  After defining and describing the notion of 
dynamical accessibility,   Kre\u{i}n-like theorems regarding signature and bifurcation are proven. 
In addition, a canonical Hamiltonian example, composed of a negative energy oscillator coupled to a 
heat bath, is treated and our development is compared to pervious work in this context.   A careful counting of eigenvalues,  with and without symmetry, is undertaken, leading to the definition of a degenerate continuum steady state (CSS) bifurcation.  It is described how the CHH and CSS bifurcations can be viewed as  linear normal forms associated with the nonlinear  single-wave model described in \cite{SW}, which is a companion to the present work  and that of \cite{chaptI}.  

\end{abstract}
 
\maketitle

\section{Introduction}

Bifurcations of unstable modes from the continuous spectrum underlie pattern formation in a wide variety of physical systems that can be described by Hamiltonian 2+1 field theories. These patterns take the form of vortices in phase space, and are referred to as `BGK modes' (Bernstein, Greene, Kruskal) in plasma physics \cite{bernstein58a} and `Kelvin cat's-eye' vortices in two-dimensional, inviscid, incompressible fluid mechanics, and possess analogs in condensed matter physics, geophysical fluid dynamics, and astrophysics. The equations that describe these physical systems all share crucial features: a formulation as a noncanonical Hamiltonian system \cite{morrison82,morrison98}, and
that stable equilibria possess continuous spectra.  Before nonlinear patterns form in these systems, unstable modes  bifurcate from their continuous spectra, a linear bifurcation we call the continuum Hamiltonian Hopf (CHH) bifurcation that is an analog   of the usual Hamiltonian Hopf (HH) bifurcation of finite-dimensional systems. In this chapter we describe some mathematical aspects of the CHH, continuing on from the material presented in \cite{chaptI}.

 Perturbation of point spectra in canonical, finite-degree-of-freedom Hamiltonian systems is described by Kre\u{i}n's theorem  \cite{krein50,krein2,moser58}, which states that a necessary condition for a HH  bifurcation is to have a collision between eigenvalues of opposite signature. A  different situation arises in the infinite-dimensional case if the linear Hamiltonian system has a continuous spectrum.  A representative example of such a Hamiltonian system is the Vlasov-Poisson equation \cite{morrison80},   which when linearized about stable homogeneous equilibria gives rise to a linear Hamiltonian system with pure continuous  spectrum that can be brought into action-angle normal form \cite{MP92,morrison94,ms94,morrison00a}.  A definition of signature was given in these works for the continuous spectrum.  The primary example here will be the Vlasov-Poisson  equation,  but the same structure is possessed by a large class of  equations \cite{morrison03}, examples being  Euler's equation for the two-dimensional fluid, where signature for shear flow
 continuous spectra was defined \cite{BM98,BM02}, and likewise for a model for  electron temperature gradient turbulence \cite{tassi1}. Modulo  technicalities, the behavior treated here is expected to cover a large class of systems.

In Sec.~\ref{sec:mathCHH} we present the mathematical structure that we use to describe CHH bifurcations, in particular we define structural stability and discuss the definition of signature for the continuous spectrum. One of the crucial parts of this framework is the choice of the norm on the perturbations to the time evolution operator, a step that requires selection of a Banach space to be the phase space for solutions  of the linearized system. In Sec.~\ref{sec:Vlasov} we apply this framework to the Vlasov-Poisson equation, presenting without proof results that  appeared  in \cite{hagstrom}. We show that the plasma two-stream instability is a CHH bifurcation that can be viewed as a zero-frequency mode interacting with a negative energy continuous spectrum to bifurcate to instability, so that the continuous spectrum provides the `other' mode in the CHH bifurcation. We show that if in the chosen Banach space the $\sup$ of the Hilbert transform is an unbounded operator, then equilibria of the Vlasov-Poisson equation are always structurally unstable. Two examples of such Banach spaces are $W^{1,1}(\R)$ and $C^0(\R)\cap L^1(\R)$.  If we restrict perturbations to those that are dynamically accessible \cite{morrison98}, which precludes the possibility of altering the signature of the continuous spectrum, we prove that equilibria with positive signature only are structurally stable.

Section \ref{sec:Canonical} contains a description of the differences between canonical and noncanonical systems; in particular, comparison to  the work of Kre\u{i}n \cite{kreindaleckii} on canonical Hamiltonian systems is made and a simple demonstration of a bifurcation to instability in such a canonical  system is  described.  In Sec.~\ref{sec:SingleWave} we present the idea that a certain mean field Hamiltonian system, the single-wave model \cite{tennyson94a,delcastillo98a,SW}, is a nonlinear normal form for the CHH bifurcation that describes the eventual  nonlinear saturation of the resulting instability near criticality.  We note, this model is derived by means of matched asymptotic expansions of a Hamiltonian 2+1 mean field theory near a marginally stable equilibrium, and also by comparison with the results of numerical simulations \cite{SW}. Finally, in Sec.~\ref{sec:conclude} we summarize and conclude.

\section{Mathematical aspects of the continuum Hamiltonian Hopf bifurcation}
\label{sec:mathCHH}

In Sec.~III.B.1 of \cite{chaptI} we presented a specific example of the CHH bifurcation, the plasma two-stream instability. This theory gives a necessary condition for structural instability: collision of eigenvalues of opposite signature. We present a framework for bifurcations in noncanonical Hamiltonian systems with continuous spectra. The key notion will
be the generalization of signature to the continuous spectrum, which is prevalent in the linear infinite-dimensional Hamiltonian systems that undergo the CHH bifurcation. We first set the stage by discussing structural stability.

\subsection{Structural stability}
\label{sssec:structstab}

Now we consider  structural stability of linear noncanonical Hamiltonian systems with continuous spectrum. The dynamical variable $\zeta$ is assumed to be a member of a function space $\calb$. We are given a Hamiltonian functional $H$ ($H_L$ of \cite{chaptI}), which is  (typically) an unbounded quadratic functional on $\calb$, and a noncanonical Poisson bracket $\{\cdot,\cdot\}$, which will be bilinear, antisymmetric, and satisfy the Jacobi identity and which in this chapter will always be of Lie-Poisson form, see \cite{morrison98,marsden}. Hamilton's equations are (see the previous chapter, \cite{chaptI}):
\bq
\zeta_t=\{\zeta,H\}=\T\zeta\,.
\eq
Here $\T$ is the time evolution operator, which by assumption is a linear operator from $\calb$ to itself. This equation can also be written:
\bq
\zeta_t=\J\frac{\delta H}{\delta\zeta}=\J \A\zeta \label{eq:noncanon}
\eq
where $\J$ is the cosymplectic operator of the bracket $\{\cdot,\cdot\}$, and $\A$ is a linear operator derived
from $H$ using the bracket. Care must be taken when using this formulation as the operator $\A$ is often not
defined as an operator on $\calb$, and only the product $\J \A=\T$ takes values in $\calb$. The process of canonization, which reformulates Eq.~(\ref{eq:noncanon}) in terms of a canonical cosymplectic operator $\J_c$, which is bounded and invertible, 
eliminates this difficulty.

The operator $\T$ (and hence the linear Hamiltonian system) is \emph{spectrally stable} if
the spectrum of $\T$ is contained in the imaginary axis, $\sigma(\T)\in i\R$. This is equivalent to the condition that the spectrum is in the closed lower half plane, i.e. $Re(\sigma(\T))\leq 0$, because the spectrum satisfies the property $\lambda\in\sigma(\T)$ implies $\bar{\lambda}\in\sigma(\T)$, a property that comes  from the Hamiltonian structure. 
Solutions of spectrally stable systems grow at most sub-exponentially.

We consider now a family of Hamiltonian systems that depend continuously on some parameter, and look for changes in the stability of the Hamiltonian system as the parameter varies. Such a family can be generated in many ways. One common scenario is for the linear Hamiltonian system to come from the linearization of some nonlinear Hamiltonian system about an equilibrium solution. In that case, the bracket and Hamiltonian functional come from linearizations of the original Hamiltonian system, and both will depend on the equilibrium $f_0$ (cf.~\cite{chaptI}). Both the bracket $\J$ and Hamiltonian $H$ are subject to change, however, and this malleability of the bracket gives the bifurcation theory of noncanonical Hamiltonian systems a different character than that of canonical Hamiltonian systems.

Bifurcations to instability occur when a spectrally
stable system becomes spectrally unstable. The following definitions depend on our definition of the size of a perturbation, and we will have to choose some set of perturbations and a measure of this size in order to proceed. Assuming that we have made this choice, we say that the spectrally stable time evolution operator $\T$ is \emph{structurally stable} if there exists some $\epsilon$ such that for all perturbations $\delta \T$ satisfying $\|\delta \T\|<\epsilon$ the operator $\T+\delta \T$ is spectrally stable, where $\|\cdot\|$ is our chosen measure for perturbations of $\T$. Otherwise we say that $\T$ is \emph{structurally unstable}.

The theory will depend on the choice of allowable perturbations and norm. Let the family of Hamiltonian systems be parametrized by a parameter $\lambda$ so that the time evolution operator for each system is $\T_\lambda$. The continuity properties of the family $\T_\lambda$ will typically come from an induced norm from the Banach space in which solutions to the equations live. For instance, one choice would be that  $\T-\T_\lambda$ is a bounded 
operator.  Other choices are relatively compact/bounded perturbations (considered in \cite{grillakis} in the context of canonical systems) or the more general class of unbounded perturbations based on the \emph{gap}, given by \cite{kato}.

Some of the most physically interesting families of systems come from linearizing a Hamiltonian system about a member of a continuous family of equilibrium states. If this is the context of our physical problem, it makes sense to only consider 
perturbations that leave this Hamiltonian structure unchanged, for instance restricting to perturbations
that change the equilibrium state only. Our most important example will be the two-stream instability described by   
the Vlasov-Poisson system, which is of this type. A further restriction is to choose to perturb to equilibria that are dynamically accessible from the original equilibrium, which restricts to perturbations that can be produced using Hamiltonian forces.

\subsection{Normal forms and signature}

The Kre\u{i}n signature is essential in the description of HH bifurcations, as the collision between positive and negative energy modes is a necessary condition for the existence of a HH  bifurcation.
It is straightforward to compute the energy signature of modes in the finite-dimensional case as one can simply use the Hamiltonian function. In the infinite-dimensional case this is complicated by the presence of the continuous spectrum. The continuous analogs of discrete eigenmodes, the so-called generalized-eigenfunctions, are distributions whose Hamiltonian is not defined, and another approach, based on the the theory of normal forms, is required to attach a signature to the continuous spectrum.

The linear theory of finite-dimensional Hamiltonian systems is organized around normal forms, and the proof of Kre\u{i}n's theorem is based on the theory of normal forms. Though the situation can be more complex in the infinite-dimensionsal case, for many important  cases it is possible to find the appropriate normal form. The simplest normal forms arise when the time evolution operator of the Hamiltonian  system is diagonalizable, in which case the Hamiltonian can be written in terms of action-angle variables $(\theta(u),J(u))$ and a canonical Poisson bracket, for instance:
\bq 
H=\int_{\Gamma} du\, \si(u) \, \omega(u) J(u)
 =\frac1{2}\int_{\Gamma} du \, \si(u)\, \omega(u) \big(Q(u)^2+P(u)^2\big)\,, 
 \label{eq:oscform}
\eq
where $\Gamma\subset\R$ and in the second equality we introduce the alternative form in terms of the canonically conjugate oscillator variables $Q(u)$ and $P(u)$.  Here $\om\in\R^{>0}$ for all $u\in \Ga$ and $\si(u)\in\{\pm 1\}$ defines the signature of the continuous spectrum corresponding to $i\omega(u)$.

The ability to define the signature for a given Hamiltonian system is directly related to the ability to bring the system into a normal form, i.e.,  to canonize and diagonalize it. Diagonalization is equivalent to finding a transformation that converts the time-evolution operator of the system into a multiplication operator, viz.,  to finding  a linear operator $\L$ such that $\L\T\L^{-1}$ is a multiplication operator. The systems described in this paper all tend to have non-normal time-evolution operators, so it may be surprising that it is ever possible to define such a transformation since the spectral theorem does not apply, but it turns out that for many  important cases the time evolution operators are diagonalizable. Operators with this property are called \emph{spectral operators}, \cite{Dunford}. By definition, a spectral operator possesses a family of spectral projection operators $\E(\delta)$ defined on Borel susbets $\delta\subset\mathbb{C}$ that commute with the time evolution operator $\T$ and reduce its spectrum, i.e.,  $\sigma(\T\E(\delta))\subset\delta$. 

The signature of the  subset $\delta$ is then defined by the sign of the Hamiltonian operator 
restricted to members of $\E(\delta)\calb$, which can be positive, negative, or indefinite. For a given point
$u\in\mathbb{C}$, this is defined by taking limits of sets $\delta$ that contain $u$. If a diagonalization is known,  the application of this definition can be straightforward. Consider Eq.~(\ref{eq:oscform}) with $\si\omega(u)=u$, $u\in \R^{>0}$. Then $\E(\delta)\calb$ is equal to the functions with support on $\R\cap(-i\delta)$ 
and the energy is clearly positive. An equivalent definition involves the sign of the operator $\J$ on the spectrum but the definition involving signature is more intuitive physically.

\section{Application to Vlasov-Poisson} 
\label{sec:Vlasov}
 
Now consider  the Vlasov-Poisson system of Sec.~III.B.1 of  \cite{chaptI}, as an example of the Hamiltonian 2+1 field theories that exhibits the CHH bifurcation. Here we are interested in the properties of the equations linearized around a homogeneous equilibrium $f_0$. These equations, repeated for completeness, are
\bq
 \ze_{k,t}= - i kp \ze_k+{if_0'}\, {k^{-1}}\!  \int_{\mathbb{R}}\!d\bar{p}\,  \ze_k(\bar{p},t)=:-\T_k  \ze_k\,.
\label{fk}
\eq
Our goal is to understand how the spectrum of $\T_k$ changes under changes in $f_0$. To this end we consider consider perturbations of the  equilibrium function, $f_0+\delta f_0$. The time
evolution operator of the perturbed system is $\T_k+\delta \T_k$, where:
\bq
\delta \T_k= - {i\delta f_0'}\, {k^{-1}}  \! \int_{\mathbb{R}}\!d\bar{p}\, \ze_k(\bar{p},t)\,. 
\eq
We use the operator norm induced by the norm on $\calb$ to measure the size of $\delta \T_k$, which requires  restriction to function  spaces in which $\delta \T_k$ is bounded, for instance the Sobolev spaces $W^{1,1}(\R)$ and $C^n(\R)\cap L^1(\R)$. The quantity $\|\delta \T_k\|$ can be bounded by  $\|\delta f_0'\|$ in  the norms for $W^{1,1}$ and  $C^n \cap L^1$, because the integral operator $\int_{\mathbb{R}}\!d\bar{p}\, \ze_k(\bar{p},t)$ is a bounded operator from those spaces to $\R$:
\bq
\|\delta \T_k\|\leq \left\|\frac{\delta f_0'}{k}\right\|\,.
\eq
If we were to consider  $L^2(\R)$, then $\delta \T_k$ would be an unbounded operator and we would have to use some other means of determining its size (see Grillakis\cite{grillakis} or Kato\cite{kato}).

 As mentioned earlier, the  linearized Vlasov-Poisson system has a continuous spectrum. Morrison \cite{morrison00a} constructed a transformation  that diagonalizes the linearized Vlasov-Poisson system, converting the time evolution operator to a multiplication operator and determining the signature of the continuous spectrum. (See \cite{morrison03} for a generalization of this method to other 2+1 Hamiltonian field theories.) This is based on the $G$-transform, which for stable equilibria $f_0$ with no discrete modes is defined as follows:
\bq
G[g]=\varepsilon_R g+\varepsilon_I \calh[g] \quad\mathrm{and}\quad 
\hat{G}[f]=\frac{\varepsilon_R}{|\varepsilon|^2}f-\frac{\varepsilon_I}{|\varepsilon|^2}\calh[f]
\eq
where $\hat{G}[f]$ operating on a function $f$ in its domain is the inverse of $G$,   
\bq
\varepsilon_I:=-\frac{\pi}{k^2}f_0' \,,\qquad \varepsilon_R:=1+\calh[\varepsilon_I] \,,\qquad |\varepsilon^2|:=\varepsilon_R^2+\varepsilon_I^2\,,
\eq
and  $\calh$, the {Hilbert Transform},  is defined as
\bq
\calh[g]= \frac{1}{\pi}\dashint_{\R}\! dp\,  \frac{g(p)}{p-u}\,,
\eq
where the expression $\dashint_{\R}$ stands for the Cauchy principal value.
If $f_k$ satisfies the linearized Vlasov-Poisson equation, then $g_k=\hat{G}[f_k]$ satisfies:
\bq
\frac{\partial g_k}{\partial t}+ikug_k=0\,, 
\eq
which gives a representation of the solution upon back-transforming.

Using this $G$-transform and the theory of generating functions, it is possible to canonize and diagonalize the linearized Vlasov-Poisson system. Canonization proceeds by transforming the Poisson bracket of Eq.~(1.35) of \cite{chaptI} according to 
\bq
\mathcal{Q}_k=\frac{1}{\sqrt{2}}\left(\ze_k+\ze_{-k}\right) \qquad\qquad \mathcal{P}_k=\frac{k}{i \sqrt{2}f_0'}\left(\ze_k-\ze_{-k}\right)\,.
\eq
The canonically conjugate pair $(\mathcal{Q}_k(p),\mathcal{P}_k(p)$ are real due to the reality of the distribution function $f=f_0 +\ze$, and the Poisson bracket is in terms of them is 
\bq
\{F,G\}=\int_{\R}dp\, \left(\frac{\delta F}{\delta \mathcal{Q}}\frac{\delta G}{\delta  \mathcal{P}}-\frac{\delta G}{\delta \mathcal{Q}}\frac{\delta F}{\delta  \mathcal{P}}\right)\,,
\eq
where here and often henceforth we suppress the $k\in\N$ dependence.  

Diagonalization is achieved using a mixed-variable generating function involving the  $G$-transform, a transformation that  was  inspired by Van Kampen's formal expression for continuum eigenmodes \cite{vankampen55a},  
\bq
\T_k\mathcal{V}(u,p)=iku\, \mathcal{V}(u,p).
\eq
where 
\bq
\mathcal{V}(u,p):=\varepsilon_I(p){PV}\frac{1}{u-v}+\varepsilon_R(p)\delta(u-p),
\eq
which clearly bares the mark of the  $G$-transform. Diagonalization proceeds from  the following mixed-variable generating functional \cite{morrison00a}:
\bq
\calf_2(\mathcal{Q},P)=\int_{\R} \!du\, P\, \hat{G}\mathcal[{Q}] \,,
\eq
which leads to a transformation to the new variables $(Q,P)$, 
\bq
Q=\frac{\delta\calf_2}{\delta P}=\hat{G}[\mathcal{Q}] 
\quad \mathrm{and} \quad 
\mathcal{P}=\frac{\delta\calf_2}{\delta \mathcal{Q}}=\hat{G^{\dagger}}[P]\,.
\eq
Under direct substitution into the Hamiltonian and making use of identities derived in \cite{MP92,morrison00a}  (see also \cite{ms07})   in terms of the new variables the Hamiltonian becomes
\bq
H[Q,P]=\frac{1}{2}\int_{\R}\!du\, \sigma(u)|ku|\left(Q^2+P^2\right)\,,
\label{VPdia}
\eq 
where  $\sigma(u)=\sgn(u\varepsilon_I)$ is the signature of the continuous spectrum with frequency $\omega=|ku|$. 

\begin{figure}[htbp]
\hspace{ .2 in}
\includegraphics[scale=.8]{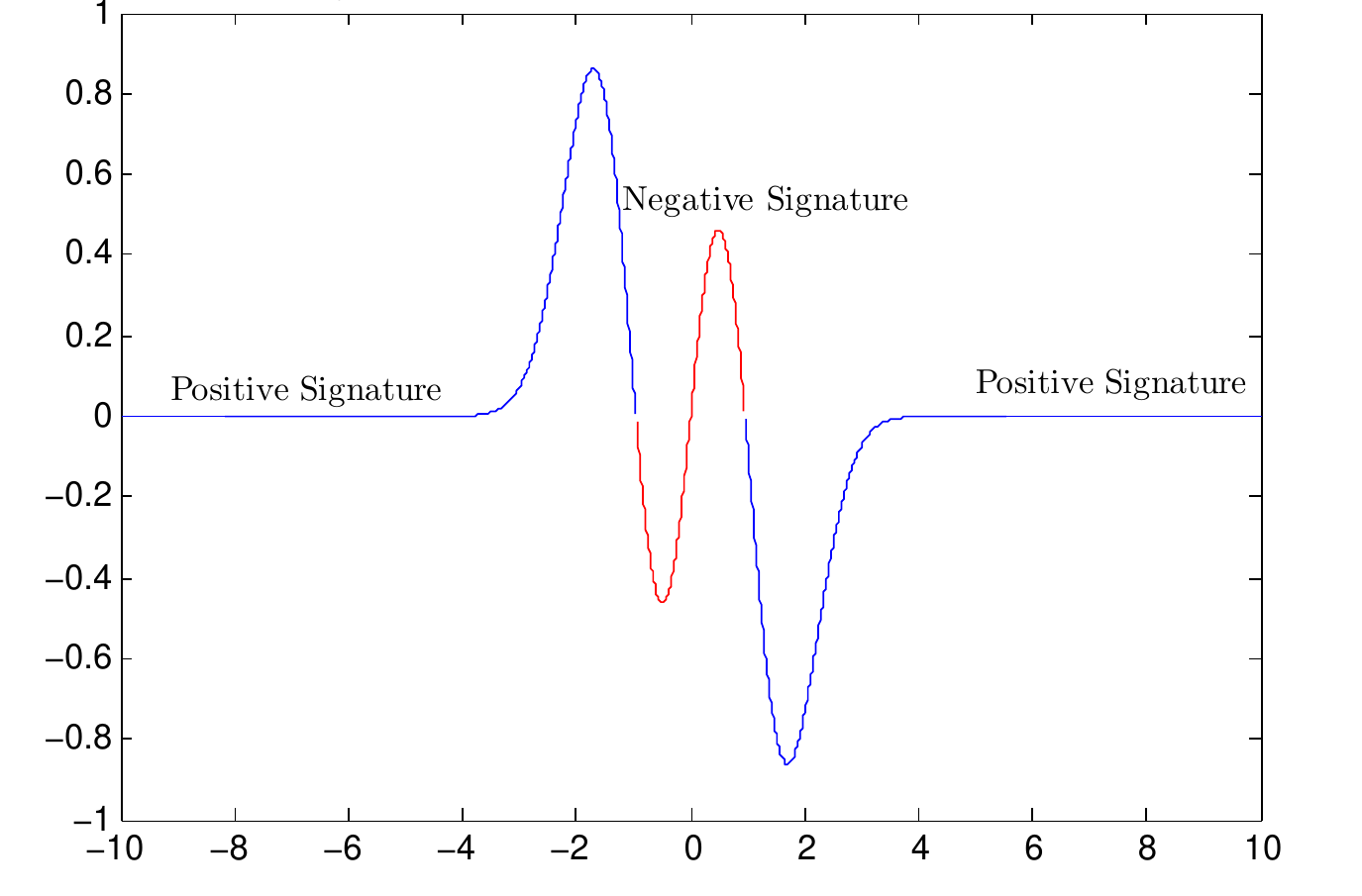}
\caption{Signature for a bi-Maxwellian distribution function.}
\label{bigaus}
\end{figure}

The Hamiltonian of (\ref{VPdia}),  that for a continuum of uncoupled harmonic oscillators,  is the normal form for the linearized Vlasov-Poisson system. This transformation can be defined only in reference frames where $f_0'(0)=0$, which can always be made true by Galilean shift.  Therefore the signature changes only when the sign of $uf_0'$ changes. 
To illustrate this signature,  consider two special cases, that of a Maxwellian distribution, $f_0=e^{-p^2}$, and that of a bi-Maxwellian distribution, $f_0=e^{-(p-p_1)^2}+e^{-(p+p_1)^2}$ (where normalization is not important). The Maxwellian distribution has one maximum and therefore it has only positive signature. On the other hand, the bi-Maxwellian has three extrema (see Fig.~\ref{bigaus}) and two signature changes.

\bigskip

The Penrose criterion, which was introduced in the previous chapter \cite{chaptI}, clearly demonstrates the role that signature plays in transitions to instability for  the linearized Vlasov-Poisson system. This criterion is that the winding number of the image of the real line under the map $\varepsilon_R+i\varepsilon_I$ is equal to $0$ when $f_0$ is stable.
Suppose that we have a family of equilibria that depend continuously on some parameter, and is  stable for some values of the parameter and unstable for others. In order for the stability to change, the Penrose plot must increase its winding number from $0$ to $1$.    For this to happen, the Penrose plot must cross the origin at the bifurcation point. We call these crossing points critical states.

A simple technique to compute the winding number is to draw a ray from the origin to infinity and
to count the number of intersections with the contour, accounting for orientation by adding $1$ for a positive orientation and subtracting $1$ for a negative orientation. 
One counts the number of zeros of
$\varepsilon_I$ for which $\varepsilon_R<0$ and adds them with a positive sign if $f_0''$ is positive,   
a crossing of the Penrose plot from the upper  half plane to the lower half plane, a negative sign if
$f_0''$ is negative, a crossing from the lower half plane to the upper
half plane, and zero if $u$ is not a crossing of the real axis, a tangency.

\subsection{Structural stability in the space $C^n(\R)\cap L^1(\R)$}

We begin by choosing the phase space of the linearized  Vlasov-Poisson system to be $C^n(\R)\cap L^1(\R)$. In this phase space, the induced norm on $\delta \T$ is proportional to the $\sup$ of $\delta f_0'$. This choice puts a restriction on the ability of perturbations to affect the signature of  $f_0'$; at a point $u$ a perturbation must have norm at least $f_0'(u)$ to induce a signature change at $u$, viz. 
\bq
\sup|\delta f_0'|\leq\|\delta \T_k\|\,.
\eq
 Furthermore, the other part of the  Penrose plot, $1-{\pi}{k^{-2}}\calh[f_0']$, is bounded in a similar way because the $\sup$ norm of the Hilbert Transform of $\delta f_0'$ is bounded by the $C^n$ norm of $\delta f_0'$, viz. 
\bq
C\sup|\calh[\delta f_0']|\leq \|\delta \T_k\|\,, 
\eq
for some constant $C$ independent of $\delta f_0'$. Therefore, for fixed $k$,  any stable $f_0$ that does not contain a discrete mode is structurally stable, as the Penrose plot will be a fixed distance from the origin. 

When $f_0$ has an embedded mode it is possible to have transitions to instability.
We identify two critical states for the Penrose plots that correspond to the transition to instability. In each of these states there is an embedded mode inside the continuous spectrum. In the first state,  the embedded mode is a so-called inflection point mode \cite{sm94}, which means there is some $\omega_c/k=u_c$ such that $\varepsilon_R(u_c)=\varepsilon_I(u_c)=0$ and $f_0''(u_c)=0$. We refer to this state as the bifurcation at $k\neq0$, because changing the value of $k$ would not cause a bifurcation. In the other state, $f_0''\neq0$,  which we call the bifurcation at $k=0$. This is named so that in a system with infinite spatial extent the unstable mode first appears at $k=0$.

\begin{figure}[htb]
  \begin{center}
    \subfigure[]{\includegraphics[scale=.55]{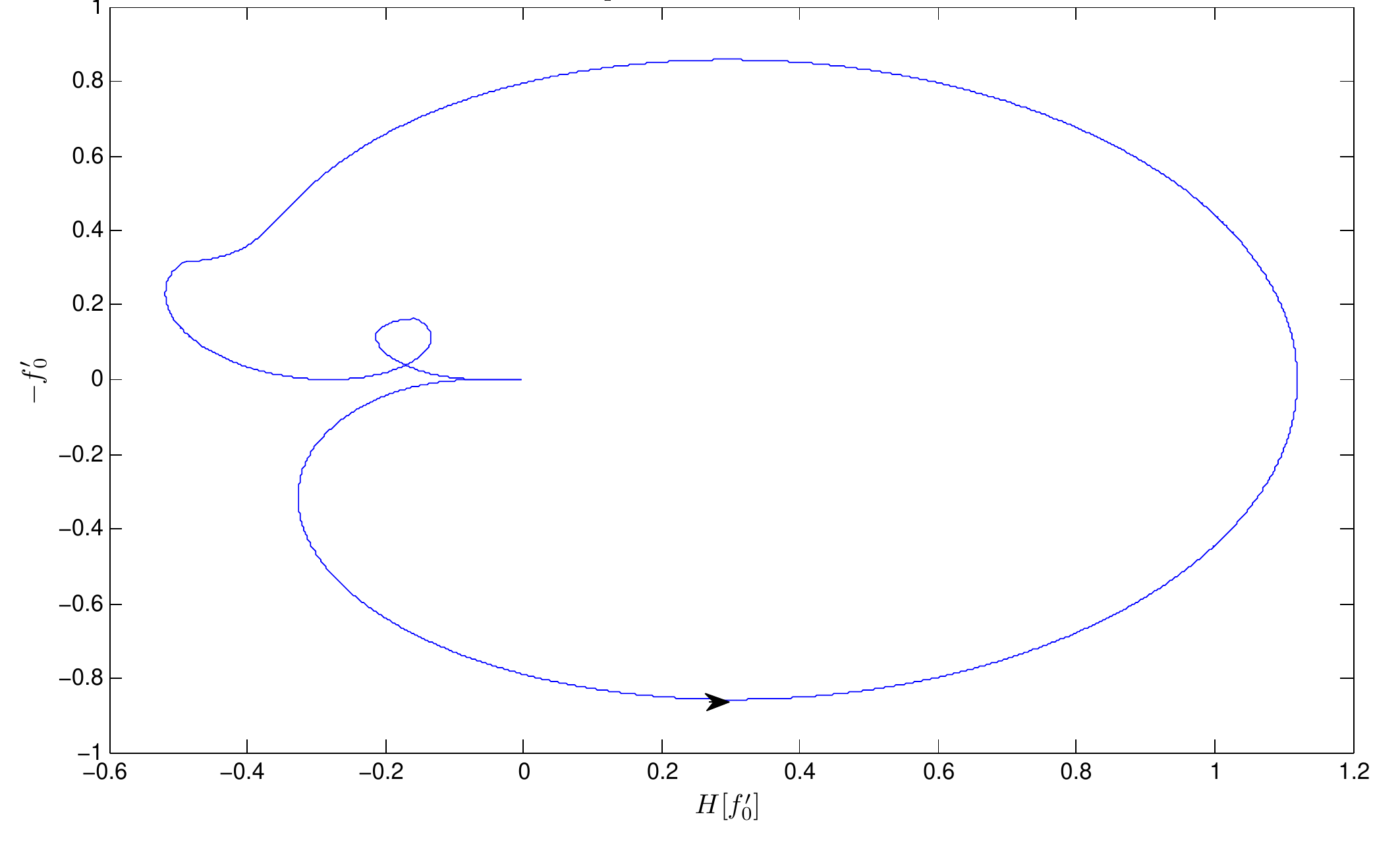}}
    \subfigure[]{\includegraphics[scale=.55]{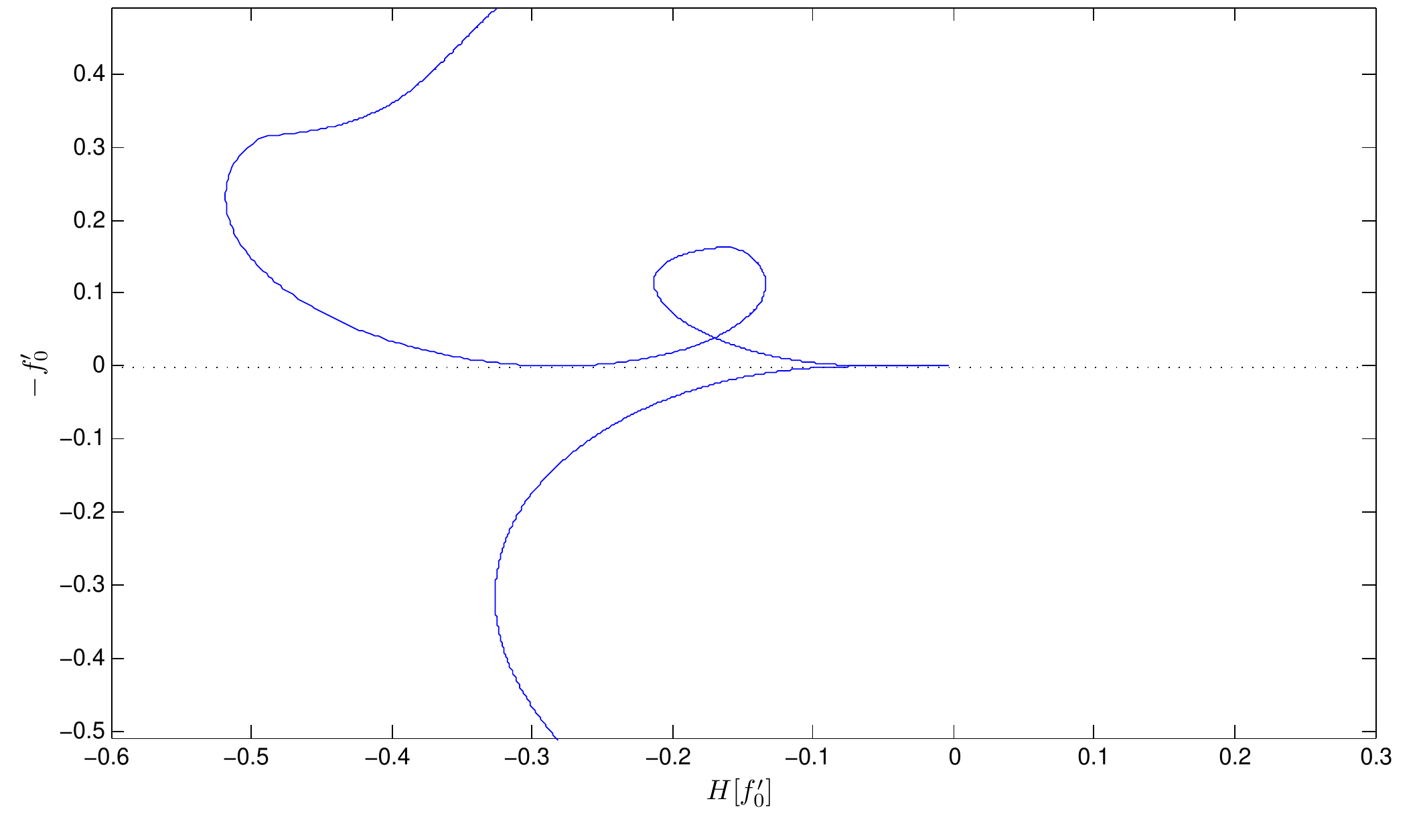}}
  \end{center}
  \caption{a) Critical Penrose plot for  a $k\neq 0$ bifurcation. b) Close up of panel a.}
\label{gausscrit}
\end{figure}

The first critical state occurs when $f_0'(u)=0$, $f_0''(u)=0$, and $1-{\pi}k^{-2}\mathcal{H}[f_0']=0$. At such a point, the addition of a generic function $\delta f_0$ to $f_0$
will cause the Penrose plot to intersect the real axis transversely, and such a perturbation can be used to cause instability. 
If the system is
perturbed so that the tangency becomes a pair of transverse intersections, then  the winding number of the Penrose plot
would jump to 1 and the system would be unstable. Figure \ref{gausscrit}  illustrates  a critical Penrose plot for a bifurcation at $k\neq 0$.  

\begin{figure}[htbp]
\begin{center}
\includegraphics[scale=.55]{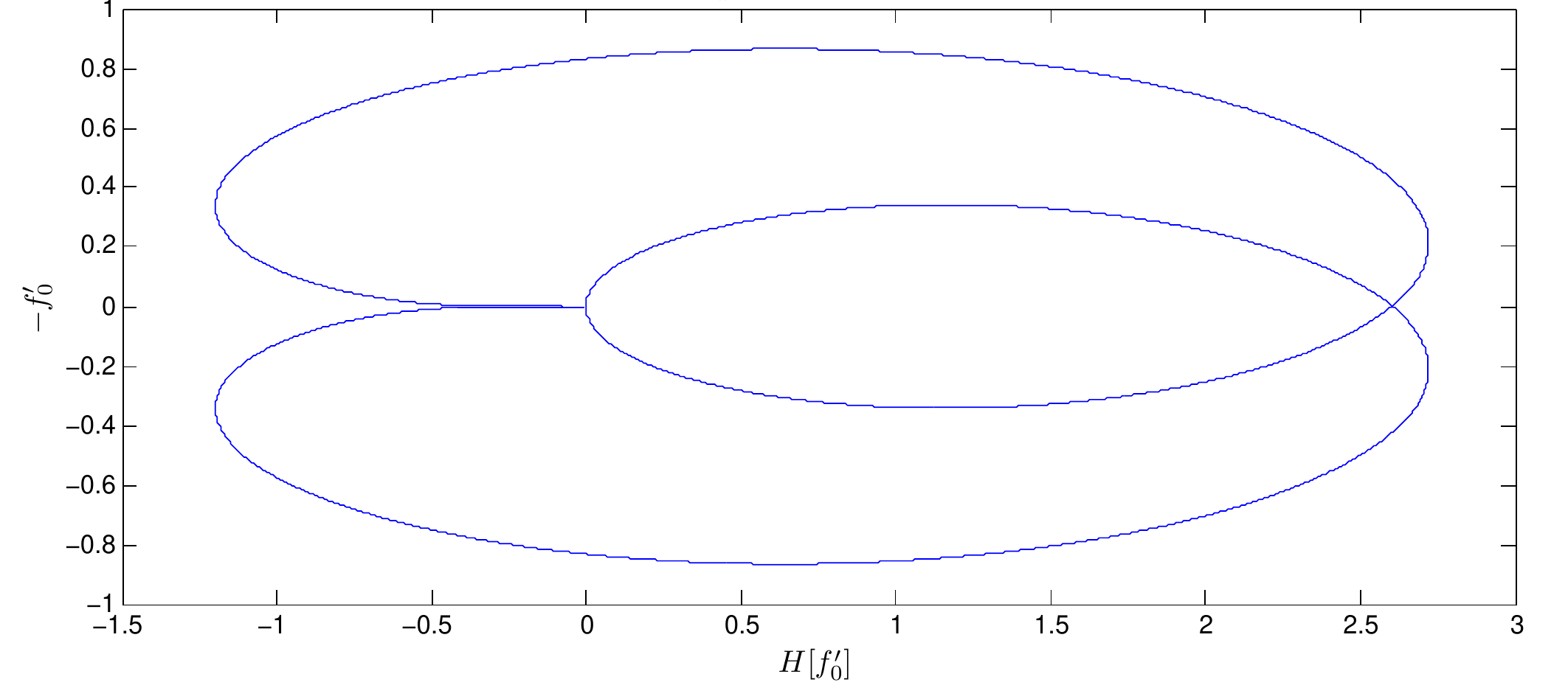}
\end{center}
\caption{Critical Penrose plot for a bi-Maxwellian distribution function.}
\label{bigaus2}
\end{figure}

Another critical state occurs when $1-{\pi}k^{-2}\calh[f_0']=0$ at a point
where $f_0'$ transversely intersects the real axis.  
If the Hilbert transform of $f_0'$ is perturbed,  there
will be a crossing with a negative $\mathcal{H}[f_0']$, and the winding number will be positive. 
Figure \ref{bigaus2} is  a critical Penrose
plot corresponding to the  bi-Maxwellian distribution with the maximum stable separation.

\bigskip

To understand and interpret these bifurcations we must understand the signature of the embedded modes at the critical state and also of the continuous spectrum. The energy signature of an embedded mode with frequency $u=\omega/k$ is given by $\sgn\left(u{\p \varepsilon_r}/{\p u}\right)$  \cite{MP92,sm94}.  

Consider the $k\neq 0$ bifurcation. Assume that the embedded mode is \emph{not} a zero frequency mode. Then we claim that if $u_c>0$ then ${\p \varepsilon_R}/{\p u}>0$ and if $u_c<0$ then ${\p \varepsilon_R}/{\p u}$<0. This is demonstrated in \cite{hagstrom} using the analyticity of the plasma dispersion function $\varepsilon$. 
Suppose the opposite were true, that $u_c>0$ and ${\p \varepsilon_R}/{\p u}<0$, then a small perturbation would generically decrease the winding number rather than increase it. This would imply the existence of poles in the upper half plane, violating the analyticity of the plasma dispersion function. This implies that the signature of the inflection point mode is always the same as the signature of the surrounding continuous spectrum. The fact that this bifurcation occurs when there is only positive signature may seem counterintuitive, but it is due to the fact that negative signature can be added in the neighborhood of the inflection point mode with an infinitesimal perturbation. When we restrict to dynamically accessible perturbations, which we do in a later section, this bifurcation will disappear.

In the $k=0$ bifurcation,  the signature of the embedded mode and the continuous spectrum surrounding it are always indefinite, either the embedded mode is at $0$ frequency, then the embedded mode has zero energy and the signature surrounding it is negative (the embedded mode is always in a valley of the distribution function, which can be seen again using analyticity and the perturbation introduced in the next section (cf.\ \cite{hagstrom})), or there is a change in the signature of the continuous spectrum. There is no reference frame in which the signature of the continuous spectrum \emph{and} embedded mode are definite.

\subsection{Structural stability in $W^{1,1}$}

We will prove that if the perturbation function is some homogeneous $\delta f_0$ and the space
is $W^{1,1}(\R)$, then  every equilibrium distribution function is structurally unstable
to an infinitesimal perturbation. Under this choice of $\calb$,  $\sup|\calh[f_0']|$ is an unbounded operator, i.e., there exists an infinitesimal $\delta f_0$ such that $\calh[\delta f_0']$ is order one at a zero of $f_0'$. Such a 
perturbation can turn any point where $f_0'=0$ into a point where $\calh[f_0'+\delta f_0']>0$
as well -- thereby  changing the winding number by moving the zeros of the Penrose plot and causing a bifurcation to instability.

We  explicitly demonstrate this structural instability for the Banach space $W^{1,1}(\R)$ and, 
by extension, the Banach space $L^1(\R)\cap C^0(\R)$, and this will imply  that every stable distribution 
function is structurally unstable, a physically unappealing state of affairs. 

\begin{figure}[htb]
  \begin{center}
    \subfigure[]{\includegraphics[scale=.557]{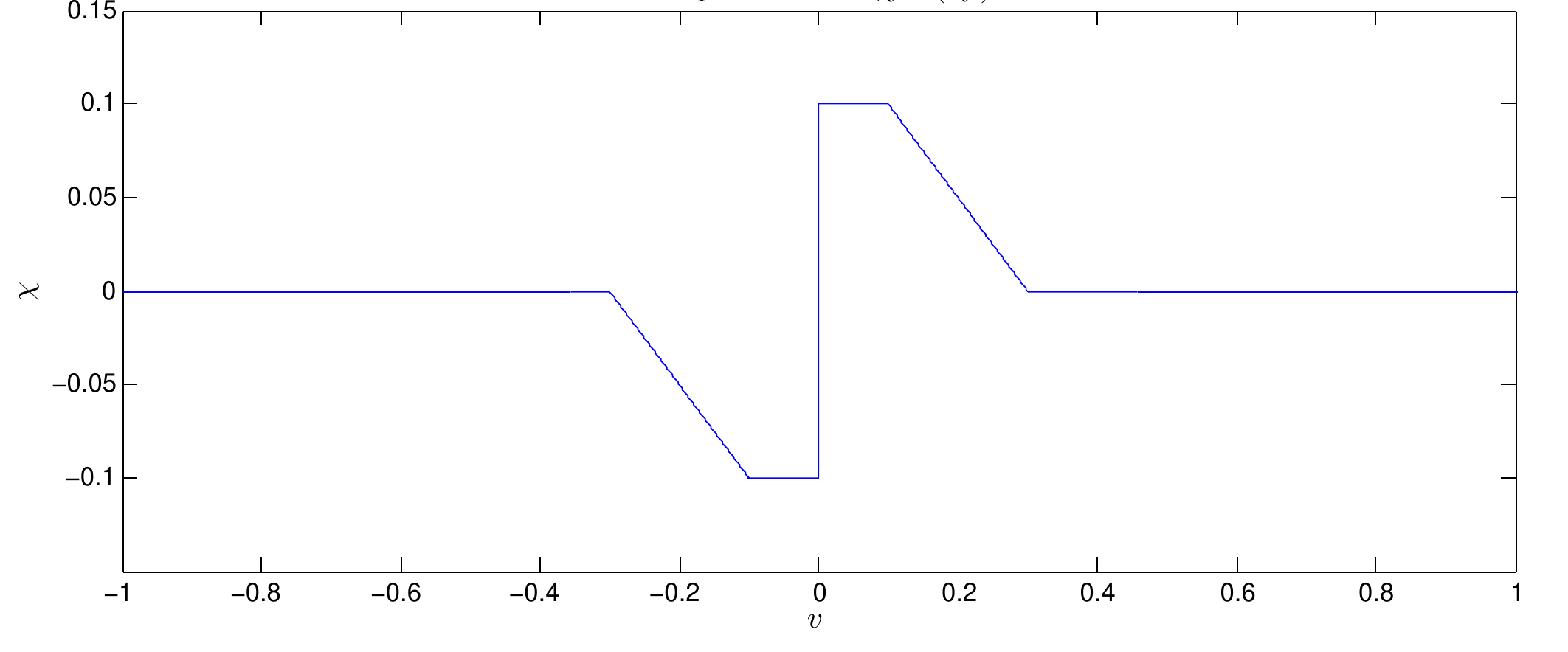}}
    \subfigure[]{\includegraphics[scale=.73]{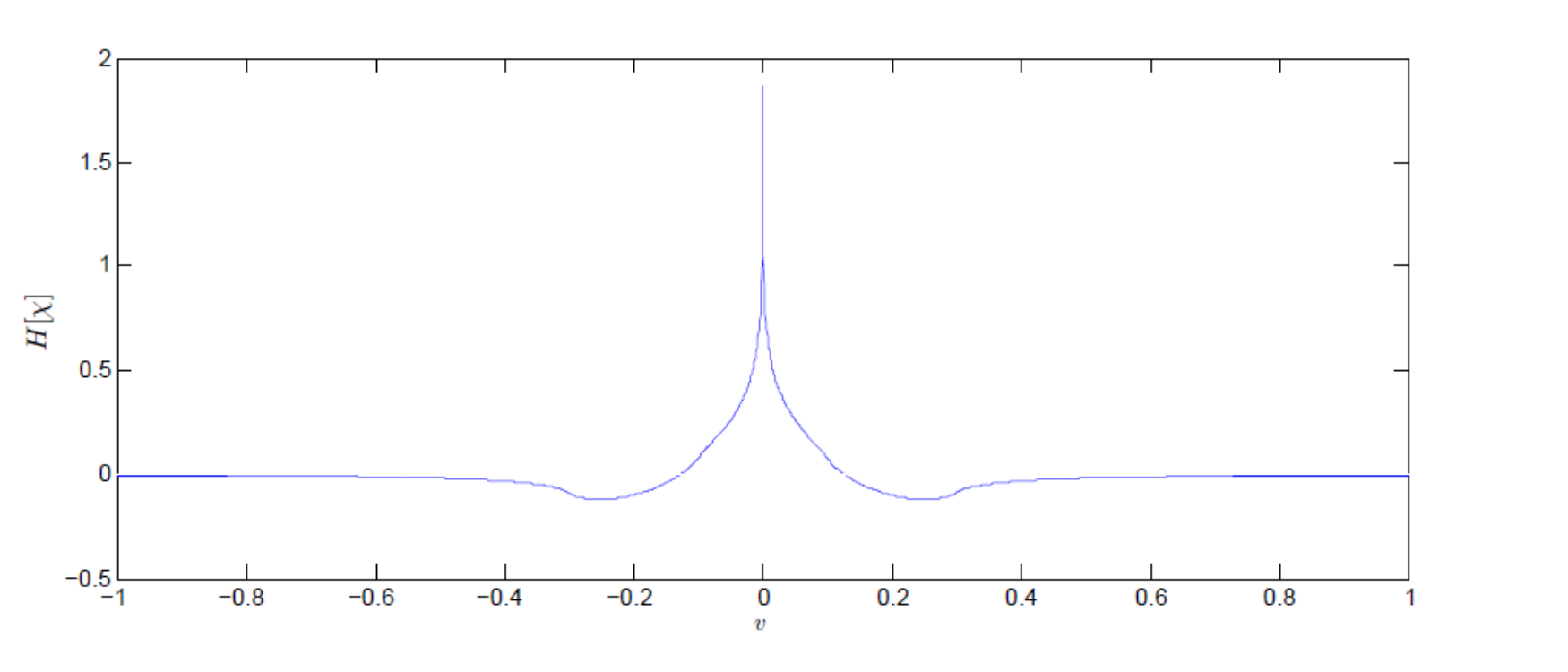}}
  \end{center}
  \caption{a) The perturbation $\chi$ for $\epsilon=e^{-10},h=d=.1$.  b) The Hilbert transform of $\chi$.}
\label{chi}
\end{figure}

Suppose we perturb $f_0$ by a function $\delta f_0$. The resulting perturbation to
the operator $\T_k$ is the operator mapping ${\ze}_k$ to $\delta f_0'\int\! dp \, {\ze}_k$. In the space
$W^{1,1}(\R)$ this is a bounded operator and thus $\|\delta f_0'\|\geq\|\delta \T_k\|$. Yet, it is possible to introduce a class of perturbations that can be  made infinitesimal,  but have Hilbert transform of order unity.   For example, consider 
the function  $\chi(p,h,d,\epsilon)$ defined by 
\[
\chi = 
\left\{
\begin{array}{ll}
{hp}/{\epsilon} &\quad  |p|<\epsilon \\
  h\, \sgn(p) &\quad \epsilon<|p|<d+\epsilon \\
  h+d/2+\epsilon/2-p/2 &\quad 2h+d+\epsilon>p>d+\epsilon \\
  -h-d/2-\epsilon/2-p/2 &\quad 2h+d+\epsilon>-p>d+\epsilon \\
  0 &\quad |p|>2h+d+\epsilon
 \end{array}\,.
 \right.
 \]
Figures \ref{chi}a and \ref{chi}b  show the graph of $\chi$ and its Hilbert transform, $\calh[\chi]$, respectively. In the space $W^{1,1}(\R)$ the function $\chi$ has norm $2h^2+2hd+h\epsilon + 4h$.  If we  choose $h=d$ and $\epsilon=O(e^{-1/h})$, then the terms that do not involve $\epsilon$ are all smaller than $(6h+\epsilon)\log (6h+\epsilon)$.  With these choices,  $\chi$ satisfies
\bqy
\chi(0)&=&2-(h+e^{-1/h})\log(|h+e^{-1/h}|)
\nonumber\\
&&
\hspace{2 cm} +(3h+e^{-1/h})\log(|3h+e^{-1/h}|)
\nonumber\\
&=&2+O(h\log h)\,.
\eqy
If we choose $d=h$ and $\epsilon=e^{-(1/h)}$, then  for any $\delta,\gamma>0$ we can choose
 an $h$ such that $\|\chi\|_{1,1}<\delta$ and $\dashint dp\, {\chi}/{p}>1-O(h)$, 
 and $|\dashint dp\, {\chi}/({u-p})|<|\gamma/u|$ for $|u|>|2h+d+\epsilon|$.

\bigskip

The perturbation $\chi$ arbitrarily moves the crossings of the real axis of the Penrose plot of $f_0$. If we use this perturbation to move crossings from the positive imaginary axis to the negative real axis, we can increase the 
winding number of the Penrose plot, thus causing instability.
Therefore the existence of this $\chi$ implies that any equilibrium is structurally  unstable in both 
the spaces $W^{1,1}(\R)$ and $L^1(\R)\cap C^0(\R)$.

\begin{theorem}
\label{structuralinstability}
A stable equilibrium distribution $f_0$ is structurally unstable under perturbations of the equilibrium in the Banach spaces $W^{1,1}(\R)$ and    $L^1(\R)\cap C^0(\R)$. 
\end{theorem}

\begin{figure}[htb]
  \begin{center}
    \subfigure[]{\includegraphics[scale=.55]{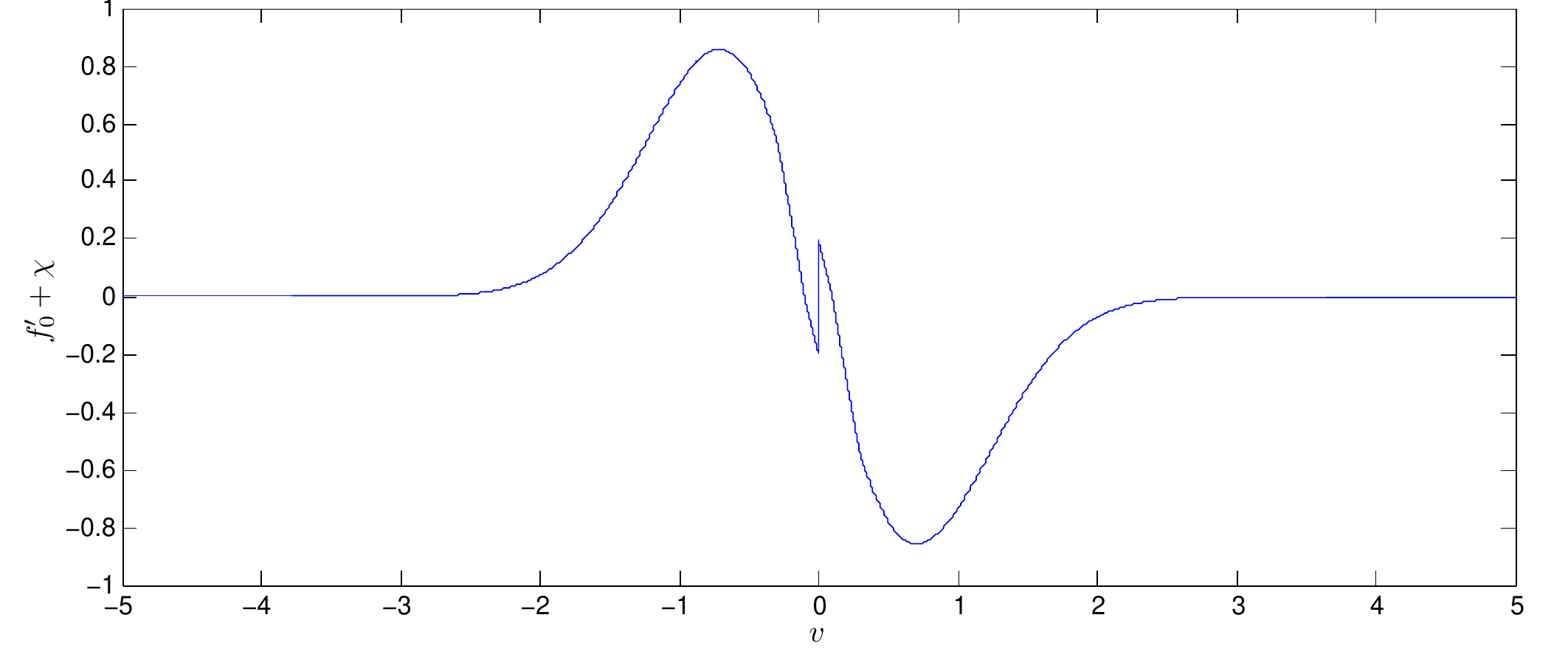}}
    \subfigure[]{\includegraphics[scale=.55]{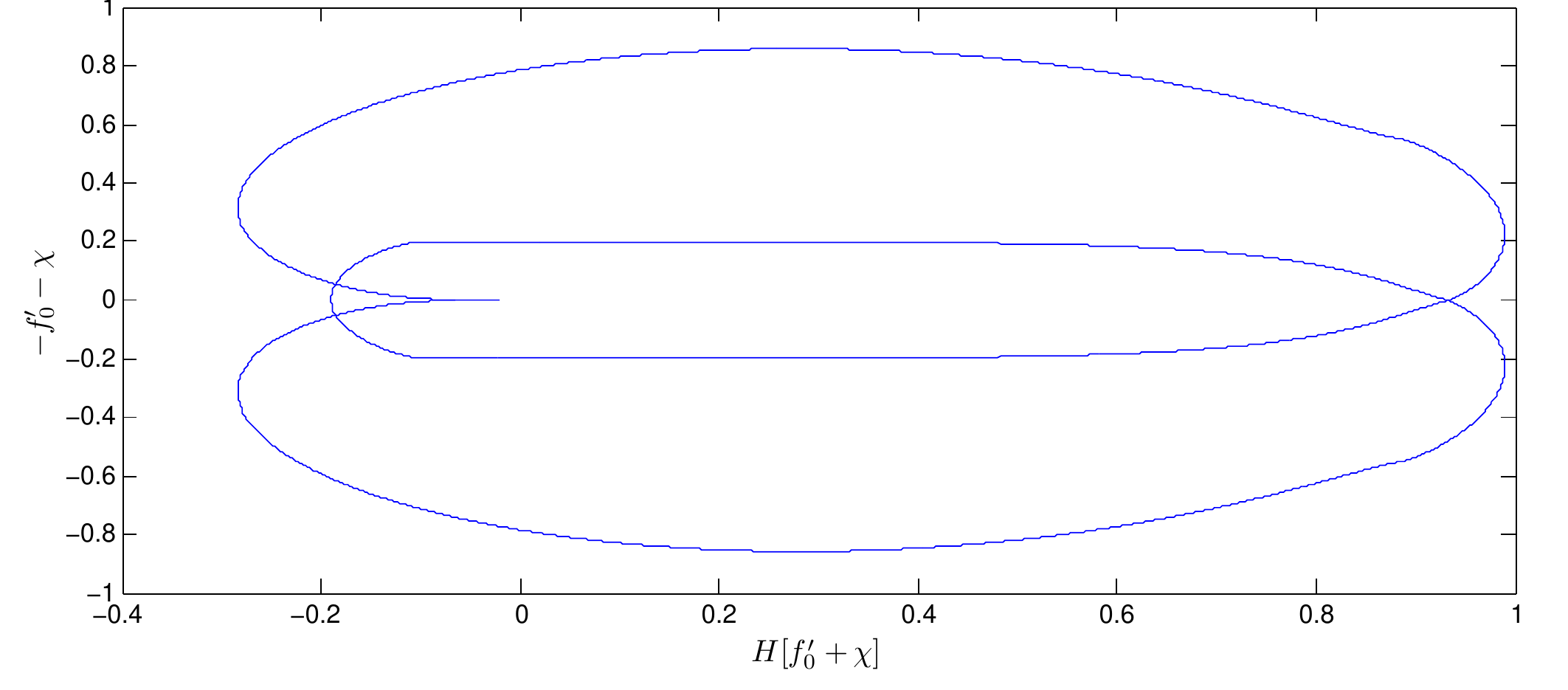}}
  \end{center}
  \caption{a) Perturbed  Maxwellian distribution, $f_0'+\chi$.  b) The Hilbert transform of panel a.}
\label{pert}
\end{figure}

Thus we emphasize that we can always construct a perturbation to $f_0$ that makes our linearized Vlasov-Poisson system unstable.    For the special case of the Maxwellian distribution,  Fig.~\ref{pert}a shows the perturbed derivative of the distribution function and  Fig.~\ref{pert}b shows the Penrose plot of the unstable perturbed system. Observe the two crossings created by the perturbation  on the positive axis as well as the negative crossing arising from the unboundedness of the perturbation. 

Theorem \ref{structuralinstability} expresses the fact that in the norm $W^{1,1}$, signature changes give rise to unstable modes under infinitesimal perturbations combined with the fact that a signature change can be induced in the neighborhood of any maximum of $f_0$.  In the next section we will demonstrate the role of signature more explicitly by restricting to  dynamically accessible perturbations. 

\subsection{Dynamical accessibility and structural stability}

As we have stated, the signature of the continuous spectrum in the Vlasov-Poisson system  is $\sgn(u\varepsilon_I(u))$. In $W^{1,1}(\R)$, $\sup|\delta f_0'|$ is bounded by $\|\delta f_0'\|$, which means that most points of the continuous spectrum cannot change signature under infinitesimal perturbations, the exception being near  points where $f_0'$ vanishes. All signature changes can be prevented by restricting to perturbations of $f_0$ that are \emph{dynamically accessible} from $f_0$, as we shall explain.    The solution of any of the mean-field Hamiltonian field
theories that have been described here can be written as a composition an initial condition $\mathring{f}$
with a symplectic map $Z(q,p)$, where $Z(q,p)$ describes the single particle characteristics (see \cite{mp90}). We say that two functions $f_1$ and $f_2$ are dynamically accessible from each other if there exists some
symplectic map $Z$ such that $f_1=f_2\circ Z$, i.e., $f_1$ is a symplectic rearrangement of $f_2$ and vice versa. 

In this work we only study perturbations of $f_0$ that preserve homogeneity. It is impossible to construct a dynamically accessible  perturbation of $f_0$ in a finite spatial domain that preserves spatial homogeneity.
To see this we write a rearrangement as $(q,p)\leftrightarrow (\hat{q},\hat{p})$,  where $\hat{p}$ is a function of
$p$ alone.  Because  $[\hat{q},\hat{p}]=1$ and  $\hat{p}(p)$ is not a function of $q$, we have 
$\hat{p}'{\p \hat{q}}/{\p q}=1$, or   $\hat{q}={q}/{\hat{p}'}$. If the spatial domain is finite,  this map is not a  diffeomorphism unless $\hat{p}'=1$.  For infinite spatial domains  this is not  a problem and these rearrangements exist. 

The reason that all homogeneous equilibria were structurally unstable in the previous section was because small perturbations could create regions of different signature near critical points of $f_0$. In fact, the Penrose criterion requires that $f_0$ has a minimum in order for there to be an unstable mode. The perturbation $\chi$ that was used to destabilize $f_0$ created a distribution function with derivative $f_0'+\chi$ that had a local minimum at what was previously a local maximum of $f_0$.  However, dynamically accessible perturbations cannot change level set topology and, consequently, the number of critical points of $f_0$.  Indeed,  if $(\hat{q},\hat{p})$ is an area preserving diffeomorphism and  $\hat{p}$  is a homogeneous, i.e., a function of $p$ alone, then the critical points of $f_0(\hat{p})$ are the points $\hat{p}^{-1}(p_c)$,  where $p_c$ is a critical point of $f_0(p)$. By the chain rule, these critical points will always be the same type as the corresponding critical point of $f_0$ --  the map $\hat{p}$ must be monotonically increasing in order for it to be invertible.

One  implication is that the perturbation $\chi$ is not dynamically accessible when it is applied to a local maximum and, consequently,   all equilibria $f_0$ with only a single critical point are structurally stable under dynamically accessible perturbations.

If $p_c$ is a nondegenerate critical point of $f_0$ such that $f_0''(p_c)<0$, then the previous obstruction to the application of $\chi$ using a dynamically accessible perturbation does not apply. In \cite{hagstrom} it is shown that there is a rearrangement $\hat{p}$ such that $f_0(\hat{p})=f_0(p)+\int_{-\infty}^pdp'\, \chi(p'-p_c)$
or that ${df_0(\hat{p})}/{dp}=f_0'(p)+\chi(p-p_c)$. Such a rearrangement can be constructed as long as the parameters defining $\chi$,
the numbers $h,d,\epsilon$,
are chosen such that $f_0'(p)+\chi(p-p_c)$ has the same critical points as $f_0'(p)$. The construction uses Morse theory to find a $\hat{p}$ so that $f_0(\hat{p})=f_0(p)+\int\chi+O((p-p_c)^3)$, where $O((p-p_c)^3)$ has compact support and is smaller than 
$f_0(p)-{f_0''(p_c)}(p-p_c)^2/2$.

These ideas lead to the following Kre\u{i}n-like theorem for dynamically accessible perturbations
in the $W^{1,1}$ norm:

\begin{theorem}
\label{kreinlike}
Let $f_0$ be a stable equilibrium distribution function for the Vlasov equation on an
infinite spatial domain. Then $f_0$ is structurally stable 
under dynamically accessible perturbations in $W^{1,1}(\R)$,  if there is only one solution 
of $f_0'(p)=0$. If there are multiple solutions,  $f_0$ is structurally unstable and the
unstable modes come from the zeros of $f_0'$ that satisfy $f_0''(p)<0$.
\end{theorem}

The implication of this result is that in a Banach space where the Hilbert transform is
an unbounded operator,  the dynamical accessibility condition makes it so that a change in
the Kre\u{i}n signature of the continuous spectrum is a necessary and sufficient condition
for structural instability. The bifurcations do not occur at all points   where the 
signature changes, however. Only those that represent valleys of the distribution can 
give birth to unstable modes. 

Dynamical accessibility also clarifies bifurcations to instability of inflection point modes. Dynamically accessible perturbations cannot eliminate inflection points of $f_0$. Since  $f_0'(u_c)$ changing sign at some point $u_c$ is necessary for instability, it is impossible for a dynamically accessible perturbation of an $f_0$ that has an inflection point mode and otherwise only a continuous spectrum with positive signature  to be unstable. This is consistent with the fact 
that there exists a frame in which signature of the continuous spectrum and the signature of the inflection point mode are both positive.

\section{Canonical infinite-dimensional case}
\label{sec:Canonical}

There have been some works on structural stability of infinite-dimensional Hamiltonian systems. The first of these results is due again to Kre\u{i}n  and recorded in his book with Dalecki\u{i} \cite{kreindaleckii}  on ordinary differential equations on Banach spaces. They considered the simplest possible infinite-dimensional Hamiltonian systems, canonical equations with bounded time evolution operators on Hilbert spaces. They defined signature in terms of positive and negative splittings of the canonical symplectic 2-form (Lagrange bracket) on the Hilbert space, the resulting condition derived in this case is the following:  if the part of the spectrum corresponding to the positive space overlaps with the part of the spectrum corresponding to the negative case, then there is an infinitesimal perturbation that causes the system to become unstable. This result applies when there is a continuous spectrum as well as a discrete spectrum, and is a direct generalization of Kre\u{i}n's finite-dimensional theorem. The splitting of the spectrum into positive and negative signature subspaces can be converted into an equivalent splitting in terms of positive and negative energy, though delicacy is again required when the spectrum is continuous. In these cases one looks at whether the Hamiltonian operator is positive or negative definite on the spectral projections onto the targeted parts of the spectrum. The slightly different definition of signature is useful when the Hamiltonian functional is allowed to depend on the time $t$, which was also studied by Kre\u{i}n, but otherwise is equivalent to our definition.

In particular, Kre\u{i}n examined \emph{canonical} equations of the form 
\bq
f_t=\J \A f \label{eq:canonical}
\eq
where $\J$ is an antisymmetric unitary operator which without loss of generality can be assumed to be $\J=\begin{pmatrix} 0 & -1\\ 1 & 0\end{pmatrix}$, and ${\A}$ is the self-adjoint operator associated with some sesquilinear form $H[\cdot,\cdot]$. Equation (\ref{eq:canonical}) is a Hamiltonian system with Hamiltonian functional $H[f,f]$. Kre\u{i}n said this system was strongly stable (structurally stable in our terminology) if there is some $\delta>0$ such that for all $|\A_1-{\A}|<\delta$, the spectrum of $\J  {\A}_1$ is contained in
the imaginary axis. Kre\u{i}n was able to prove that the system was strongly stable if and only if the phase space $\calb$ splits into two subspaces, $\calb_+$ and $\calb_-$, each invariant under the time evolution operator $\J {\A}$, such that $\J$ is positive on $\calb_+$ and negative
on $\calb_-$. This is equivalent to the Hamiltonian operator ${\A}$ being positive on $\calb_+$ and negative on $\calb_-$, which means that the system is structurally stable as long as positive energy parts of the spectrum are disjoint from the negative energy parts of the spectrum. No reference is made to the type of spectrum of the operator $\J {\A}$, and the sign of the operator ${\A}$ on the eigenspace corresponding to some part of the spectrum defines the signature of that part of the spectrum. 


The situation is more complicated when ${\A}$ is allowed to be unbounded. This case was considered by Grillakis \cite{grillakis}, who was interested in studying the stability of travelling waves in the nonlinear Schrodinger (NLS) equation and other similar systems. He was also interested in developing a technique for determining the number of negative eigenvalues, a problem subsequently treated by a number of authors \cite{chugunova1,kapitula1}. In the case where there was a negative energy mode embedded in the continuous spectrum (which had positive signature in those examples), Grillakis was able to prove structural instability. In the case where all signatures were positive, under the assumption of \emph{relatively bounded} perturbations, Grillakis was able to prove structural stability. It should be noted that in the NLS case the nature of the continuous spectrum is different than in the case of Vlasov-Poisson and the other continuous media field theories that exhibit CHH bifurcations.  In the NLS equation it is  due to the action of a derivative operator on a function space over an unbounded domain rather than a multiplication operator. In the last section of this paper we will argue that the nonlinear evolution and saturation of the resulting instability of Vlasov-Poisson and similar equations is described by something called the single-wave model. It would be interesting to see if there is an analog of the single wave model for systems like NLS, at least in some sense. This would be related to the greater issue of how the two types of continuous spectra are related.

\subsection{Negative energy oscillator coupled to a heat bath}

An illustrative example of the CHH bifurcation in the canonical case comes from a negative energy version of the Caldeira-Leggett model. This case is like that for the noncanonical equations considered  in the bulk of this chapter because the continuous spectrum arises from a multiplication operator. The Caldeira-Leggett model is a simple model of a discrete mode embedded into a continuous spectrum. It is used to introduce dissipation into quantum mechanics \cite{caldeira,hagstromCL} through the process of phase mixing, essentially realizing  the  phenomenon of  Landau damping in quantum mechanics.  (Landau damping is a symptom of the continuous spectrum, which leads to highly oscillatory solutions whose moments  decay with time as determined by the Riemann-Lebesgue lemma.) 
By flipping the sign of the signature of the discrete oscillator, we alter the Caldeira-Leggett model to describe a gyroscopically stabilized system interacting with a heat bath (see also Bloch et al.~\cite{bloch}). This  results in 
 structural instability, where the small parameter is the amplitude of the coupling term.  We demonstrate this result through an adaptation of the Nyquist method, resulting in a Penrose-like criterion for stability. 

The Hamiltonian for this system is 
\bqy
H[Q,P,q(x),p(x)]&=&-\frac{\Omega}{2}\big(Q^2+P^2\big)+\frac1{2}\int_0^{\infty} \!\!dx\, x  \big(q(x)^2+p(x)^2\big)
\nonumber\\
&& \hspace{1.5 cm}+ Q \int_0^{\infty} \!\!dx\, f(x)q(x) \,, 
\eqy
If $f(x)=0$, the Hamiltonian describes a system consisting of a single harmonic oscillator with negative energy and a continuous bath of oscillators with positive energy,  where $(q(x),p(x))$ are coordinates for  the bath and here $(Q,P)$ the single harmonic oscillator. Solutions are stable and consist of independent oscillations of the discrete oscillators and the continuum. If the discrete oscillator has positive energy, and we activate the coupling to the continuum, then because the  spectrum is always of positive signature,  we will still have stable solutions. In the negative signature case, we expect the opposite.   This can be seen by an argument that is analogous to the Penrose criterion in the Vlasov equation.

The equations of motion are 
\bqy
\frac{dQ}{dt}=-\Omega P \qquad\qquad \frac{dP}{dt}=\Omega Q -\int_0^{\infty}\! \!dx\,  f(x)q(x)\\
\frac{\partial q(x)}{\partial t}=xp(x) \qquad\qquad \frac{\partial p(x)}{\partial t}=-xq(x)-Qf(x)\,, 
\eqy
which have the dispersion relation
\bq
(\Omega^2-\omega^2)=\Omega \int_0^{\infty}\!\! dx\,  \frac{xf(x)^2}{\omega^2-x^2}\,.
\label{cldr}
\eq
Here we use partial fractions to write the integral on the right hand side of (\ref{cldr}) in terms of the Cauchy integral of the anti-symmetric extenstion of $f(x)$, denoted by $f_-(x)$, 
\bq
(\Omega^2-\omega^2)-\Omega \int_{\R}\! dx\,  \frac{f(x)_-^2}{2(\omega-x)}\, .
\eq
If we divide both sides by $\omega^2+\Omega^2$ and take the limit as $\omega$ approaches the real axis from the upper half plane, we  get the following expression for the dispersion relation on the real axis:
\bq
\varepsilon(\omega)=\frac{\omega^2-\Omega^2}{\omega^2+\Omega^2}-\frac{\Omega}{\omega^2+\Omega^2}\frac{\pi}{2} \calh[f(x)^2_-](\omega)-\frac{i\pi}{2(\omega^2+\Omega^2)}f(\omega)^2_-\,. 
\label{CLdr2}
\eq

\begin{figure}[htbp]
\begin{center}
\includegraphics[scale=.57]{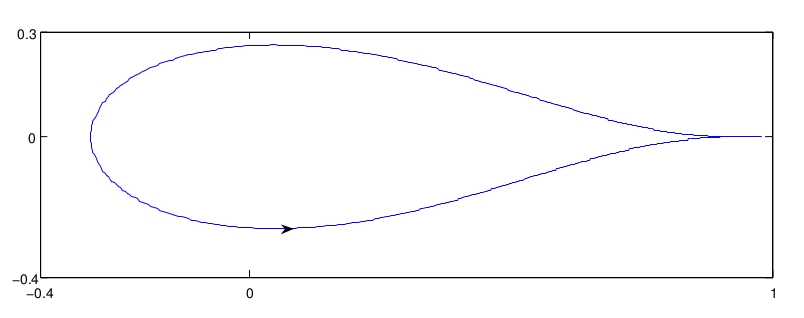}
\end{center}
\caption{Nyquist plot for the Caldeira-Leggett model with a negative energy harmonic oscillator at frequency $\Omega=1.0$ and coupling function $f(x)^2=.4xe^{-.25x^2}$.}
\label{CLPPenrose}
\end{figure}

Using the argument principle, we find that the number of zeros in the upper half plane is equal to the winding number of the image of the real line under this mapping minus the number of poles. Since there is a single pole, where $\omega=i\Omega$, the number of zeros is the winding number plus 1. For $x>0$, the imaginary part of the image is negative;  for $x<0$,  it is positive. For generic $f$ (not too large),  the winding number of this contour will be $1$, and there will be two zeros of the dispersion relation in the upper half plane, see Fig.~\ref{CLPPenrose} for such an example. These zeros emerge due
to an interaction between the continuous spectrum and the discrete mode with opposite signature, just like in the CHH  bifurcations that we have discussed so far.

\section{Commentary:  degeneracy and nonlinearity}
\label{sec:SingleWave}

We have given arguments above and in \cite{chaptI} that the CHH is like the usual HH except the continuous spectrum plays the role of one of the colliding eigenvalue pairs in the discrete case.  A telltale that a continuous spectrum is playing this role is the presence of an imaginary part to the dispersion relation when evaluated at real frequencies, as is the case, e.g.,  for the Vlasov-Poisson system where on the real $u$ axis $\vep(k,u):=1-\pi k^{-2}  \mathcal{H}[f_0'](u)\pm i \pi k^{-2} f_0'(u)$.  The two signs occur because the real axis is a branch cut,  which is known to be a consequence of continuous spectra in systems of this type. Observe, the same occurs for the Caldeira-Leggett model in (\ref{CLdr2}).  After collision, the number of discrete eigenvalues the emerge can be counted in a  straightforward way.  For example,  consider the Penrose plot of Fig.~\ref{gausscrit} that depicts a $k\neq 0$ bifurcation  at criticality. If the $f_0(p)$ used for Fig.~\ref{gausscrit} is replaced by $f_{\eta}(p)$, a one parameter perturbation that matches $f_0(p)$ at $\eta=0$, then  when instability sets in the point of tangency will move with $\eta$ so that there are two intersections of the real axis giving rise to a winding number of unity, which signals the instability.  Generically this will give a complex eigenvalue where $\om= \om_R + i \ga$, with both real and imaginary parts of $\om$ depending on the bifurcation parameter $\eta$.   A similar Penrose argument reveals that there is also a root in the lower half plane, bringing our eigenvalue count to two, with $\om= \om_R - i \ga$ corresponding to decay.  In these plots  $k$ is assumed to be fixed, but associated with a given $k\in \N$ is canonical pair,  $(\mathcal{Q}_k,\mathcal{P}_k)$,  which can be traced back to  $\ze_k$ and $\ze_{-k}$.  Here each $k\in \N$ labels a degree of freedom,  which has two associated  eigenvalues:  a mode or degree of freedom, determined by its wavenumber,  has two dimensions,    corresponding  to  amplitude and phase.  Replacing $k$ by $-k$ in $\varep$ gives the remaining two eigenvalues, $\om= -\om_R \pm i \ga$.  Thus,  the CHH is a bifurcation to a quartet, $\om= \pm\om_R \pm i \ga$, and  after bifurcation the structure is identical to that of the ordinary HH bifurcation. 

Tractability often arises in problems because of assumptions of symmetry, e.g., the homogeneity of the equilibrium $f_0$ simplifies the Vlasov problem  and the symmetry in the Jeans problem of Sec.~II.B.3 of \cite{chaptI} allowed an explicit solution of the dispersion relation (25).  Thus, the question arises, what happens if we symmetrize the $k\neq 0$ CHH bifurcation discussed above?   If $f_0(p)=f_0(-p)$, with the upper portion of  Fig.~\ref{gausscrit}  unchanged,  then we obtain a plot that is reflection symmetric about the  $\mathcal{H}$-axis with two osculating points.  Under parameter change to instability, both curves must cross and using  the ray counting procedure discussed in the in Sec.~\ref{sec:Vlasov}  this causes the winding number to jump by 2.  Thus, for symmetric $f_0$ with  $k\neq 0$,  bifurcating eigenvalues occur in pairs and after bifurcation we have an octet, characteristic of a degenerate CHH. 

Next consider the $k=0$ bifurcation with the imposed symmetry $f_{\eta}(p)=f_{\eta}(-p)$ for all control parameter values $\eta\in \R^{\geq 0}$ with  criticality at $\eta=0$ as depicted in Fig.~\ref{bigaus2}.   Because of the symmetry $f_{\eta}(0)=0$ for all $\eta$ near $\eta=0$. The bifurcation can be instigated either by fixing $k$ and varying $\eta$ or  by setting $\eta$ to a value for which the crossing of  Fig.~\ref{bigaus2}  becomes  negative and then varying $k$ until $\varep=0$.  Either way, it follows  that with the imposition of this symmetry the solution of the dispersion relation, $\varep=0$, must have the following form:
\bq
\om^2= G(k,\eta)
\label{ssbif}
\eq
where the function $G$ is real.  This is  seen by  separating  the dispersion  relation into real and imaginary parts,
\bqy
\varep(k,\om)&=&1+\frac1{k^2}\int_{\R}\! dp\, \frac{f_{\eta}'}{u-p}
\label{split}\\
&=&1+\frac{1}{k^2}\int_{\R}\!dp\, \frac{f_{\eta}' \, (u_R-p)}{(u_R-p)^2+u_I^2} 
-i \, \frac{u_I}{k^2}\int_{\R}\!dp\,\frac{f_{\eta}'}{(u_R-p)^2+u_I^2}\,,
\nonumber
\eqy
where  $u=\omega/k=u_R+iu_I$.  Then, with the  assumption that $f_{\eta}'$ is antisymmetric in its argument $p$ and splitting the imaginary part of (\ref{split}) into symmetric and anti-symmetric parts yields
\bqy
&& \hspace{-.5 cm} u_I\int_{\R}\!dp\, \frac{f_{\eta}'}{(u_R-p)^2+u_I^2} 
 = \frac{u_I}{2}\int_{\R}\!dp\, 
\frac{f_{\eta}'\left[(u_R+p)^2 -(u_R-p)^2\right]}{\left[(u_R-p)^2+u_I^2\right]\left[(u_R+p)^2+u_I^2\right]}
\nonumber\\
&& \hspace{1 cm} =2u_I u_R\int_{\R}\!dp \,
\frac{p \,f_{\eta}'}{\left[(u_R-p)^2+u_I^2\right]\big[(u_R+p)^2+u_I^2\big]}\,.
\label{integral}
\eqy
This expression must vanish when $u$ is a root, which implies that  $u_I=0$,  $u_R=0$, or the integral equals zero. 
 If we assume that $u_I$ is nonzero, then   either  $u_R$ vanishes or  the integral of (\ref{integral}) vanishes. In general, even with the assumed symmetry,  we do not  expect the integral to vanish;  the condition for the existence of an embedded mode does not reference this integral in any way, and the imaginary part of the dispersion relation at criticality for such,  only depends on the value of $f_{\eta}$ at the frequency of the mode.  Therefore,  at the bifurcation point this integral does not appear.

Note, the case of the degenerate octet discussed above is not forbidden by this argument,  due to the potential vanishing of the integral, which would allow for both $u_I$ and $u_R$ to be nonzero. From such a state, further variation of $f_{\eta}$ will lead to a branch of solutions in the upper half plane. Also note that for the Vlasov-Jeans instability,  where the sign of the interaction is reversed, the integral (\ref{integral}) has a positive integrand for the Maxwellian distribution and therefore cannot vanish. 

 
From the above discussion about symmetry, it is clear that at criticality, say at $\eta=0$,  $G(k,0)=0$ implies discrete zero frequency eigenvalues, while as $\eta$ increases $G(k,\eta>0)<0$ implies two pure imaginary eigenvalues,  indicating  exponential growth and decay.   In fact, the situation is precisely like the dispersion relation of (22)  for the multi-fluid example of \cite{chaptI}. Upon properly counting eigenvalues as above, we see that after the bifurcation there are in fact two growing and two decaying eigenvalues.  We note that an  attempt to use the usual marginality relations for determining the eigenvalues, 
\bq
\varep_R(k,\om_R)=0
\quad \mathrm{and}\quad 
\ga=-\frac{\varep_I(k,\om_R)}{\p \varep_R/\p \om_R}\,, 
\label{margin}
\eq
at $\om_R=0$, will be indeterminant because both the numerator and denominator of $\ga$ vanish.   As we have seen in \cite{chaptI}  such degenerate steady state (SS) bifurcations happen in finite systems when symmetry is imposed.  We call  any SS bifurcation in the presence of a continuous spectrum,  a continuum steady state bifurcation (CSS).   

If one breaks the symmetry, then generically as $\eta$ increases the $k=0$ bifurcation is a CHH bifurcation.  For this case  generally $f'_{\eta}$ does not vanish and equations (\ref{margin}) apply.  Counting eigenvalues gives the CHH quartet.  One might be fooled into thinking a change of frames, a doppler shift,  would make the symmetric and non symmetric $k=0$ bifucations identical, but this is not the case.  Galilean frame shifting the degenerate CSS, say by a   speed $v^*$,  replaces (\ref{ssbif}) by  a dispersion relation of the form $(\om -kv^*)^2= G(k,\eta)$;  thus, unlike the  nonsymmetric case the  real parts of the frequencies do not  depend on $\eta$.

A goal of linearized theories is to predict weakly nonlinear  behavior.  Indeed, bifurcation theory in dissipative systems has achieved great success in this regard.  In particular,  for  finite-dimensional  systems rigorous center manifold theorems allow one to reliably track bifurcated solutions into the nonlinear regime and, in some instances, obtain saturated values.  For infinite-dimensional systems various normal forms, such as the  Ginzburg Landau equation adequately describe pattern formation due to the appearance of a single mode of instability in a wide variety of dissipative problems.  In Hamiltonian systems the  situation is more complicated; the lack of dissipation  creates a greater challenge because dimensional reduction is not so accommodating.  However,  for finite-dimensional Hamiltonian systems there is a long history of perturbation/ averaging techniques for near integrable systems, systems with adiabatic invariants, etc. Techniques that may lead to nonlinear normal forms.  Similarly, techniques have been developed for infinite-dimensional Hamiltonian systems, particularly  in the context of single field 1+1 models.  However, the combination of nonlinearity together with the type of continuous spectrum treated here and in \cite{chaptI} provide  a distinctively  more difficult  challenge.

This challenge is met by the  single-wave (SW) model, an infinite-dimensional  Hamiltonian system   that   describes the behavior near threshold and subsequent nonlinear evolution of  a discrete mode that emerges from the continuous spectrum.  The SW model was originally derived in plasma physics, then (re)discovered in various fields of inquiry, ranging from fluid  mechanics, galactic dynamics, and  condensed matter physics.  The presence of the continuous spectrum, which  is responsible for Landau damping on the linear level,  causes conventional perturbation analyses to fail because of singularities that occur at all orders of perturbation. However, in  \cite{SW},  it was shown by  a suitable matched asymptotic analysis, how the single-wave model emerges from the large class of 2+1 theories of Sec.~III of \cite{chaptI}.  An essential ingredient for this asymptotic  reduction is that these Hamiltonian systems have a continuous  spectrum in the linear stability problem, arising not from an infinite spatial domain but from singular resonances along curves in phase space  (e.g., the wave-particle resonances in the plasma problem or critical levels in fluid mechanics).  Thus, the SW model describes  nonlinear consequences of the CHH and CSS bifurcations. 

\begin{figure}[htbp]
\begin{center}
\includegraphics[scale=.6]{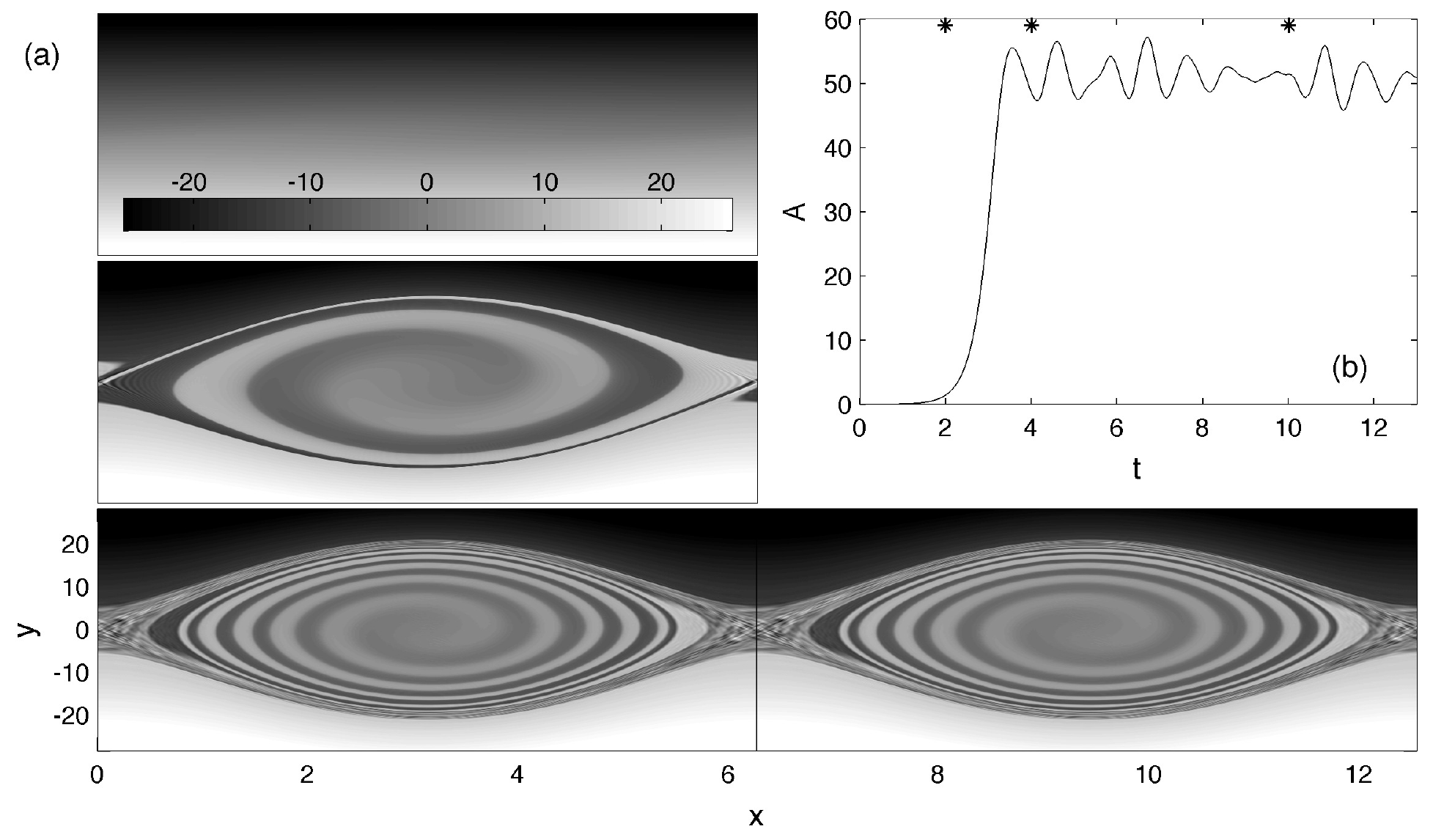}
\end{center}
\caption{Evolution of the single-wave model ({\it after Balmforth et al. Rev.~Mod.~Phys. (2013)}. a) Plot of a typical phase space hole pattern.  b)  Behavior of mode amplitude as function of time.}
\label{swm}
\end{figure}

In particular, the  SW  model describes a range  of universal phenomena,  some of which have been rediscovered in different contexts.  For a bifurcation to instability, the model features the so-called  trapping scaling dictating the saturation amplitude, and the cats-eye or phase space hole  structures that characterize the resulting phase-space patterns.  An example of this is shown in Fig.~\ref{swm}, which depicts the phase space pattern and temporal fate of the singe-wave (bifurcated mode) amplitude.  The SW model also gives a description of nonlinear Landau damping, i.e.,  how such damping can be arrested by nonlinearity.  An in-depth description of he SW model  is beyond the scope of the present contribution, but we comment that in addition to the normal form that aligns with the CHH bifurcation there is also a degenerate from associated with the CSS bifurcation.  We refer the reader to  \cite{SW} (notably Sec.~V) for further details.

\bigskip

\section{Summary and conclusions}
\label{sec:conclude}

We presented a mathematical account of CHH bifurcations in 2+1 Hamiltonian continuous media field theories. 
We presented a mathematical framework in which we describe the structural stability of equilibria of Hamiltonian systems, whose most important ingredient was a method for attaching signature 
to the continuous spectrum. We presented an application of this framework to the Vlasov-Poisson equation,
demonstrating that the two-stream instability can be interpreted as a positive energy mode interacting with 
a negative energy continuous spectrum, and that all equilibria are structurally unstable in Banach spaces that are not strong enough to prevent infinitesimal perturbations from altering the signature of the continuous spectrum. If we restrict to dynamically accessible perturbations, which by construction cannot effect the signature, then only those equilibria with both positive and negative signature are structurally unstable.

In the last parts of this paper we examined the difference between canonical and noncanonical Hamiltonian systems and also the idea that the single-wave model is a normal (degenerate) form for the CHH (CSS) bifurcation that describes it's nonlinear evolution. These processes underlie phase space pattern formation in the 2+1 theories that we are interested in, and explaining these patterns and the structure of the phase space of these systems provides strong motivation for this work and further research.

\bigskip
\noindent \textbf{Acknowledgements.}
 Hospitality of the GFD Summer Program, held at the Woods Hole Oceanographic Institution, is greatly appreciated.  GIH and PJM were supported by USDOE grant nos.~DE-FG02-ER53223 and DE-FG02-04ER54742, respectively.

\end{document}